\documentclass[twocolumn,aps,showpacs]{revtex4}
\usepackage[utf8]{inputenc}
\usepackage{color}
\usepackage{amsmath}
\usepackage{amssymb}
\usepackage{graphicx}

\makeatletter
\@ifundefined{textcolor}{}
{%
 \definecolor{BLACK}{gray}{0}
 \definecolor{WHITE}{gray}{1}
 \definecolor{RED}{rgb}{1,0,0}
 \definecolor{GREEN}{rgb}{0,1,0}
 \definecolor{BLUE}{rgb}{0,0,1}
 \definecolor{CYAN}{cmyk}{1,0,0,0}
 \definecolor{MAGENTA}{cmyk}{0,1,0,0}
 \definecolor{YELLOW}{cmyk}{0,0,1,0}
}

\usepackage{dcolumn}
\usepackage{bm}
\usepackage{wrapfig}
\usepackage{subfigure}
\usepackage{ucs}
\usepackage{lineno}
\usepackage{epsfig}
\usepackage{psfrag}\usepackage{epstopdf}
\DeclareGraphicsExtensions{.pdf,.eps,.png,.jpg,.mps}

\makeatother

\begin{document}
\title{Thermal quantum correlations of a single electron in a double quantum dot
with transverse magnetic field}
\author{Vinicius Leitão$^{1}$, Onofre Rojas$^{1}$, Moises Rojas$^{1}$}
\affiliation{$^{1}$Departament of Physics, Institute of Natural Science, Federal University of Lavras, 37200-900, Lavras-MG, Brazil}
\begin{abstract}
In this paper, we investigate the thermal quantum correlations in a semiconductor double quantum dot system. The device comprises a single electron in a double quantum dot subjected to a longitudinal magnetic field and a transverse magnetic field gradient. The thermal entanglement of the single electron is driven by the charge and spin qubits. Utilizing the density matrix formalism, we derive analytical expressions for thermal concurrence and correlated coherence. The main goal of this work is to provide a good understanding of the effects of temperature and various parameters on quantum coherence. Additionally, our findings indicate that the transverse magnetic field can be employed to adjust the thermal entanglement and quantum coherence of the system. We also highlight the roles of thermal entanglement and correlated coherence in generating quantum correlations,  noting that thermal correlated coherence is consistently more robust than thermal entanglement.  This suggests  that quantum algorithms based solely on correlated coherence might be more resilient than those relying on entanglement.
\end{abstract}
\maketitle

\section{introduction}

Quantum resource theories have emerged as a relevant field of research in 
recent years \cite{chi,stre1}. In particular, quantum coherence and quantum 
entanglement are two essential properties of non-classical systems. They can be analyzed through
an operational resource theory, aiming at applications in quantum technologies and
quantum information process \cite{benn-1,Ben2,lamico}
and emerging fields such as quantum metrology \cite{fro,gio}, quantum
thermodynamics \cite{brandao,lan} and quantum biology \cite{lambert}.
Furthermore, over the past decade, the manipulation and generation
of quantum correlations have been extensively investigated in various quantum
systems, including Heisenberg models \cite{arne,kam,ro,ro-1,carva-1,carva-2}, trapped
ions \cite{tur}, and cavity quantum electrodynamics \cite{rai,davi}
and so on.

Solid-state quantum dots (QDs) are among the most promising physical systems for implementing quantum technologies \cite{peta,shin,piotr,ladd}. There are proposals for QDs devices that utilize
either charge \cite{haya,gor,kr,kay} or spin \cite{loss} are
qubits, or even both simultaneously \cite{benito,an,mielke,yang}. These quantum
systems are of great interest due to their easy integration with
existing electronics and scalability advantage \cite{ita,urda}. Moreover,
the quantum dynamics and entanglement of two electrons within coupled double quantum dots (DQDs) have been addressed \cite{sanz,sza}, while aspects related to the quantum correlations and decoherence have been investigated \cite{fan,qin,borge,sou}. Additional properties have also been studied, including quantum teleportation based on the double quantum dots  \cite{choo}, quantum noise induced by phonons in double quantum dots charge qubits \cite{gia}, multielectron quantum dots \cite{rao}, and thermal quantum correlations in coupled double semiconductor charge qubits \cite{moi,moises,dah}. More recently, a conceptual design of quantum heat
machines has been developed using two coupled double quantum-dot systems
as a working substance \cite{moi-1}.

Quantum coherence, arising from the principle of quantum superposition, is a central concept in quantum mechanics, playing a crucial role in both bipartite and multipartite quantum correlations. The quantification of coherence relies on the definition of an incoherence basis, where incoherent states are those represented as diagonal in that basis. Over time, various methods have been development to quantify quantum coherence. Among them, those proposed by Baumgratz et al.\cite{baum}, such as the $l_1$-norm of the coherence and the relative entropy of coherence.\cite{Hu,stre2}. More recently, a new approach has emerged with the introduction of correlated coherence \cite{tan,tri}, a measure designed to investigate the relationship between quantum coherence and quantum correlations. This measure eliminates local coherence components, focusing exclusively on the coherence attributed to quantum correlations. As such, correlation coherence provides a powerful tool to understand how coherence is distributed and shared among different subsystems, offering valuable insights into the interplay between coherence and entanglement in complex quantum systems. 

In this work, we aim to investigate the role of thermal entanglement and quantum correlated coherence in a single electron spin within a double quantum dot system under the influence of an external transverse magnetic field. The gradient of the magnetic field induces spin-orbit interaction, which facilities the operation of a flopping-mode spin qubit, enabling efficient manipulation of spin states through charge dynamics. The system is assumed to be isolated from its electronic reservoirs, which are maintained in the strong Coulomb blockade regime, ensuring that only a single electron occupies the DQD. Analytical solutions were obtained, enabling a detailed exploration of the thermal evolution of the system's populations and the behavior of the thermal fidelity and derive an analytical expression for quantum entanglement. These solutions also allowed us to study the model's fidelity and derive an analytical expression for quantum correlated coherence. Additionally, we compared thermal 
entanglement with quantum correlated coherence, highlighting their relationship. Note, the framework provided by correlated coherence 
offers a unified perspective on quantum correlations, encompassing concepts such as quantum discord and quantum entanglement \cite{yue,ma}.

Our paper is organized as follows. In Section II, introduces the physical
model and outlines the method used to analyze it. Section III, provides a concise overview of the
definitions of concurrence ($\mathcal{C}$) and the correlated
coherence ($\mathcal{C}_{cc}$), followed by the derivation of the analytical expressions. In Section IV presents a 
detailed discussion the most interesting
results, including entanglement, populations dynamics, and correlated coherence. Finally, Section V is devoted to the conclusions.


\section{The model and Hamiltonian}

\begin{figure}
\includegraphics[scale=0.95]{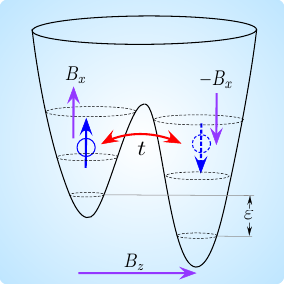}
\caption{\label{fig:Fig-1} Schematic representation of the double quantum
dot (DQD) system occupied by a single electron. The energy levels of the quantum dots are detuned by an amount $\varepsilon$. The electron's spin is depicted as a small blue sphere, delocalized between two quantum dots.}
\end{figure}

As depicted in Fig. \ref{fig:Fig-1}, we consider a silicon device
consisting of a double quantum dot filled with a single electron, which
has two charge configurations: the electron located either
on the left ($L$) or right ($R$) dot, corresponding to position
states labeled by $\left|L\right>$ and $\left|R\right>$ respectively.
The Hamiltonian of the double quantum dot \cite{benito,an} in the presence
of a homogeneous magnetic field $B_{z}$ and a magnetic field gradient
$B_{x}$ perpendicular to $B_{z}$ is given by
\begin{equation}
\begin{array}{ccc}
H & = & \frac{\varepsilon}{2}\tau_{z}+t\tau_{x}+\frac{B_{z}}{2}\sigma_{z}+\frac{B_{x}}{2}\tau_{z}\sigma_{x},\end{array}\label{eq:1}
\end{equation}
where $\tau_{x,z}$ and $\sigma_{x,z}$ are the
Pauli matrices in the position and electron spin space, respectively,
$t$ is the tunneling parameter, and $\varepsilon$ is the inter-dot detuning. The single electronic spin states are described by $\{\left|0\right>,\left|1\right>\}$.

In the standard bases $\left\{ \left|L0\right>,\left|L1\right>,\left|R0\right>,\left|R1\right>\right\} $,
the eigenvalues of the Hamiltonian $H$ are given by
\begin{eqnarray}
E_{1,2} & = & \pm\frac{1}{2}\sqrt{\Sigma+2\sqrt{\Omega}},\\
E_{3,4} & = & \pm\frac{1}{2}\sqrt{\Sigma-2\sqrt{\Omega}},
\end{eqnarray}
where $\Omega=4B^{2}_{z}t^{2}+\varepsilon^{2}(B_{z}^{2}+B_{x}^{2}),\Sigma=B_{z}^{2}+B_{x}^{2}+4t^{2}+\varepsilon^{2}$. The corresponding eigenvectors are given by
\begin{eqnarray}
\left|\varphi_{1,2}\right> & = & M_{\pm}\left[a_{\pm}\left|L0\right>+b_{\pm}\left|L1\right>+\left|R0\right>+c_{\pm}\left|R1\right>\right],\nonumber \\
\left|\varphi_{3,4}\right> & = & N_{\pm}\left[\widetilde{a}_{\pm}\left|L0\right>+\widetilde{b}_{\pm}\left|L1\right>+\left|R0\right>+\widetilde{c}_{\pm}\left|R1\right>\right],
\end{eqnarray}
where 
$M_{\pm}=\frac{1}{\sqrt{a_{\pm}^{2}+b_{\pm}^{2}+c_{\pm}^{2}+1}}$ and $N_{\pm}=\frac{1}{\sqrt{\widetilde{a}_{\pm}^{2}+\widetilde{b}_{\pm}^{2}+\widetilde{c}_{\pm}^{2}+1}}$ with 
\begin{eqnarray}
a_{\pm}&=&\frac{(\varepsilon\pm E_{1})^{2}-E_{3}^{2}}{2t(B_{z}+\varepsilon)},\\ 
b_{\pm}&=&\frac{E_{1}(\mp B_{z}\varepsilon+(\varepsilon-B_{z})E_{1}\pm(E_{1}^{2}-E_{3}^{2}))}{B_{x}t(B_{z}+\varepsilon)}\\
&&+\frac{\alpha^{2}}{4B_{x}t},\\ 
c_{\pm}&=&\frac{(B_{z}\mp E_{1})^{2}-E_{3}^{2}}{(B_{z}+\varepsilon)B_{x}},\\ 
\widetilde{a}_{\pm}&=&\frac{(\varepsilon\pm E_{3})^{2}-E_{1}^{2}}{2t(B_{z}+\varepsilon)},\\ 
\widetilde{b}_{\pm}&=&\frac{E_{3}(\mp B_{z}\varepsilon+(\varepsilon-B_{z})E_{3}\mp(E_{1}^{2}-E_{3}^{2}))}{B_{x}t(B_{z}+\varepsilon)}\\
&&+\frac{\alpha^{2}}{4B_{x}t},\\ 
\widetilde{c}_{\pm}&=&\frac{(B_{z}\mp E_{3})^{2}-E_{1}^{2}}{(B_{z}+\varepsilon)B_{x}}.
\end{eqnarray}
And $\alpha^{2}=B_{z}^{2}+B_{x}^{2}-\varepsilon^{2}-4t^{2}.$

The system state in the thermal equilibrium is described by 
$\rho(T)=\frac{\exp(-\beta H)}{Z}$, where $\beta=1/k_{B}T$, 
with $k_{B}$ being the Boltzmann's constant,
$T$ the absolute temperature, and the partition function of the
system is defined by $Z={\sum_{i=1}^{4}}e^{-\beta E_{i}}$.


\subsection{The density operator}


The thermal density operator for the double quantum dot model of a single electron in a transverse magnetic field at temperature $T$ is given by
\begin{equation}
\rho_{AB}(T)=\left[\begin{array}{cccc}
\rho_{11} & \rho_{12} & \rho_{13} & \rho_{14}\\
\rho_{12} & \rho_{22} & \rho_{23} & \rho_{24}\\
\rho_{13} & \rho_{23} & \rho_{33} & \rho_{34}\\
\rho_{14}& \rho_{24} & \rho_{34} & \rho_{44}
\end{array}\right].\label{eq:5}
\end{equation}
The elements of this density matrix, after a cumbersome algebraic
manipulation, are given by

\[
\begin{array}{ccl}
\rho_{11} & = & \frac{M_{+}^{2}a_{+}^{2}e^{-\beta\varepsilon_{1}}+M_{-}^{2}a_{-}^{2}e^{-\beta\varepsilon_{2}}+N_{+}^{2}\widetilde{a}^{2}_{+}e^{-\beta\varepsilon_{3}}+N_{-}^{2}\widetilde{a}_{-}^{2}e^{-\beta\varepsilon_{4}}}{Z},\\
\rho_{12} & = & \frac{M_{+}^{2}a_{+}b_{+}e^{-\beta\varepsilon_{1}}+M_{-}^{2}a_{-}b_{-}e^{-\beta\varepsilon_{2}}}{Z}\\
& &+\frac{N_{+}^{2}\widetilde{a}_{+}\widetilde{b}_{+}e^{-\beta\varepsilon_{3}}+N_{-}^{2}\widetilde{a}_{-}\widetilde{b}_{-}e^{-\beta\varepsilon_{4}}}{Z},\\
\rho_{13} & = & \frac{M_{+}^{2}a_{+}e^{-\beta\varepsilon_{1}}+M_{-}^{2}a_{-}e^{-\beta\varepsilon_{2}}+N_{+}^{2}\widetilde{a}_{+}e^{-\beta\varepsilon_{3}}+N_{-}^{2}\widetilde{a}_{-}e^{-\beta\varepsilon_{4}}}{Z},\\
\rho_{14} & = & \frac{M_{+}^{2}a_{+}c_{+}e^{-\beta\varepsilon_{1}}+M_{-}^{2}a_{-}c_{-}e^{-\beta\varepsilon_{2}}}{Z}\\
&&+\frac{N_{+}^{2}\widetilde{a}_{+}\widetilde{b}_{+}e^{-\beta\varepsilon_{3}}+N_{-}^{2}\widetilde{a}_{-}\widetilde{b}_{-}e^{-\beta\varepsilon_{4}}}{Z},\\
\rho_{22} & = & \frac{M_{+}^{2}b_{+}^{2}e^{-\beta\varepsilon_{1}}+M_{-}^{2}b_{-}^{2}e^{-\beta\varepsilon_{2}}+N_{+}^{2}\widetilde{b}_{+}^{2}e^{-\beta\varepsilon_{3}}+N_{-}^{2}\widetilde{b}_{-}^{2}e^{-\beta\varepsilon_{4}}}{Z},\\
\rho_{23} & = & \frac{M_{+}^{2}b_{+}e^{-\beta\varepsilon_{1}}+M_{-}^{2}b_{-}e^{-\beta\varepsilon_{2}}+N_{+}^{2}\widetilde{b}_{+}e^{-\beta\varepsilon_{3}}+N_{-}^{2}\widetilde{b}_{-}e^{-\beta\varepsilon_{4}}}{Z},\\
\rho_{24} & = & \frac{M_{+}^{2}b_{+}c_{+}e^{-\beta\varepsilon_{1}}+M_{-}^{2}b_{-}c_{-}e^{-\beta\varepsilon_{2}}}{Z}\\
& &+\frac{N_{+}^{2}\widetilde{b}_{+}\widetilde{c}_{+}e^{-\beta\varepsilon_{3}}+N_{-}^{2}\widetilde{b}_{-}\widetilde{c}_{-}e^{-\beta\varepsilon_{4}}}{Z},\\
\rho_{33} & = & \frac{M_{+}^{2}e^{-\beta\varepsilon_{1}}+M_{-}^{2}e^{-\beta\varepsilon_{2}}+N_{+}^{2}e^{-\beta\varepsilon_{3}}+N_{-}^{2}e^{-\beta\varepsilon_{4}}}{Z},\\
\rho_{34} & = & \frac{M_{+}^{2}c_{+}e^{-\beta\varepsilon_{1}}+M_{-}^{2}c_{-}e^{-\beta\varepsilon_{2}}+N_{+}^{2}\widetilde{c}_{+}e^{-\beta\varepsilon_{3}}+N_{-}^{2}\widetilde{c}_{-}e^{-\beta\varepsilon_{4}}}{Z},\\
\rho_{44} & = & \frac{M_{+}^{2}c_{+}^{2}e^{-\beta\varepsilon_{1}}+M_{-}^{2}c_{-}^{2}e^{-\beta\varepsilon_{2}}+N_{+}^{2}\widetilde{c}_{+}^{2}e^{-\beta\varepsilon_{3}}+N_{-}^{2}\widetilde{c}_{-}^{2}e^{-\beta\varepsilon_{4}}}{Z}.
\end{array}
\]

Since $\rho_{AB}(T)$ represents a thermal state in equilibrium, the
corresponding entanglement is referred to as \textit{thermal entanglement}.
In this paper, we focus on a single electron spin
in a double quantum dot subject to a transverse magnetic field. We show that the
thermal quantum coherence of the model is governed by the charge qubit, controlled by the inter-dot tunneling, and the spin qubit,
influenced by both the parallel and transverse magnetic fields.


\section{Quantum Correlations}

This section provides a concise review of the definition and
properties of thermal entanglement and quantum coherence.

\subsection{Thermal entanglement }

To quantify the amount of entanglement in a 
given two-qubit state $\rho$, we use the concurrence $\mathcal{C}$ as 
defined by Wootters \cite{wootters,woo} 
\begin{eqnarray}
\mathcal{C}={\rm {max}\left\{ 0,2max\left(\sqrt{\lambda_{i}}\right)-\sum_{i}\sqrt{\lambda_{i}}\right\}.}
\end{eqnarray}
Here, $\lambda_{i}\:(i=1,2,3,4)$ are the eigenvalues of the matrix arranged in descending
order
\begin{eqnarray}
R=\rho\left(\sigma^{y}\otimes\sigma^{y}\right)\rho^{\ast}\left(\sigma^{y}\otimes\sigma^{y}\right),
\end{eqnarray}
with $\sigma^{y}$ being the Pauli matrix. After straightforward calculations,
the eigenvalues of the matrix $R$ are given by
\begin{eqnarray}
\lambda_{1,2} & = & \Theta+G\pm\sqrt{\Xi_{+}\Sigma_{+}},\nonumber \\
\lambda_{3,4} & = & \Theta-G\pm\sqrt{\Xi_{+}\Sigma_{+}},
\end{eqnarray}
where 
\[
\begin{array}{ccl}
G & = & -2\rho_{14}\rho_{12}+\rho_{11}\rho_{24}-\rho_{13}\rho_{22},\\
\Theta & = & \rho_{11}\rho_{22}-\rho_{13}\rho_{24}+|\rho_{14}|^{2}+|\rho_{12}|^{2},\\
\Xi_{\pm} & = & 2\left(\rho_{12}\pm\rho_{14}\right)\left(\rho_{22}\pm\rho_{24}\right),\\
\Sigma_{\pm} & = & 2\left(\rho_{13}\mp\rho_{11}\right)\left(\rho_{14}\pm\rho_{12}\right).
\end{array}
\]
Thus the concurrence of this system can be written
as \cite{cao} 
\begin{eqnarray}
\mathcal{C}={\rm {max}\left\{ 0,\mid\sqrt{\lambda_{1}}-\sqrt{\lambda_{3}}\mid-\sqrt{\lambda_{2}}-\sqrt{\lambda_{4}}\right\} .}
\end{eqnarray}
In this case, the analytical expression for the thermal concurrence
is too lengthy to be provided explicitly here, but it can be easily derived by
following the steps outlined above.

\subsection{Thermal fidelity}
The mixed-state fidelity, as defined in reference \cite{jo,zh}, is given by
\begin{equation}
F(\rho_{1},\rho_{2})=Tr\left (\sqrt{\rho_{2}^{1/2}\rho_{1}\rho_{2}^{1/2}}\right ).
\end{equation}
This metric quantifies how distinguishable two quantum states $\rho_{1}$ and $\rho_{2}$. In contrast, the fidelity 
between an input pure state and output mixed state is defined by
\begin{equation}
F=\langle \psi|\rho|\psi\rangle,
\end{equation}
where $|\psi\rangle$ represents the pure state and $\rho$ the density operator of the mixed state. This latter measure informs about the overlap between the pure state $|\psi\rangle$ and the mixed state $\rho$. In our study, we examine the thermal fidelity between the ground state $|\varphi_{2}\rangle$ and the system's state at temperature $T$. After some algebraic manipulation, we find 
\begin{eqnarray}
F(T) & = & N_{-}^{2}[\widetilde{a}_{-}^{2}\rho_{11}+\widetilde{b}_{-}^{2}\rho_{22}+\rho_{33}+\widetilde{c}_{-}^{2}\rho_{44}\nonumber \\
 &  &+2\widetilde{a}_{-}(\widetilde{b}_{-}\rho_{12}+\rho_{13}+\widetilde{c}_{-}\rho_{14})\nonumber \\
 &  &+2\widetilde{b}_{-}(\rho_{23}+\widetilde{c}_{-}\rho_{24})+2\widetilde{c}_{-}\rho_{34}].
\end{eqnarray}

\subsection{Correlated coherence}

Quantum coherence is a crucial aspect of quantum physics, and it
holds practical significance in quantum information processing task.
In a bipartite system,  quantum coherence can exist both locally and in 
the correlations between the subsystems. The amount of coherence 
present in the global state,  after subtracting the purely local coherences,  is known 
as correlated coherence $\mathcal{C}_{cc}$ \cite{tan}. For a bipartite quantum
system $\rho_{AB}$, it becomes 
\begin{equation}
\mathcal{C}_{cc}(\rho_{AB})=\mathcal{C}_{l_{1}}(\rho_{AB})-\mathcal{C}_{l_{1}}(\rho_{A})-\mathcal{C}_{l_{1}}(\rho_{B}),\label{eq:6}
\end{equation}
where $\rho_{A}=Tr_{B}(\rho_{AB})$ and $\rho_{B}=Tr_{A}(\rho_{AB})$.
Here, $A$ and $B$ stand for local subsystems.

Several quantum coherence measures have been proposed to 
conform to the set of properties that any suitable coherence measure
should satisfy \cite{baum,shu}. In this regard,  the $l_{1}$-norm, has gained 
considerable attention as a reliable measure of coherence. 
The $l_{1}$-norm of coherence,  denoted as $\mathcal{C}_{l_{1}}$, is 
defined as follows

\begin{equation}
\mathcal{C}_{l_{1}}(\rho)=\sum_{i\neq j}|\langle i|\rho|j\rangle|.
\end{equation}

Quantum coherence is a concept that depends on the choice of basis. However, we can select
an incoherent basis for local coherence, allowing us to
diagonalize the density matrices $\rho_{A}$ and $\rho_{B}$. According to Eq.(\ref{eq:5}), the
reduced density matrix $\rho_{A}(T)$ can be expressed as follows
\begin{equation}
\rho_{A}(T)=\left(\begin{array}{cc}
\rho_{11}+\rho_{22} & \rho_{13}+\rho_{24}\\
\rho_{13}+\rho_{24} & \rho_{33}+\rho_{44}
\end{array}\right).\label{eq:rha}
\end{equation}
In a similar way, we obtain 
\begin{equation}
\rho_{B}(T)=\left(\begin{array}{cc}
\rho_{11}+\rho_{33} & \rho_{12}+\rho_{34}\\
\rho_{12}+\rho_{34} & \rho_{22}+\rho_{44}
\end{array}\right).\label{eq:rhb}
\end{equation}
To analyze the correlated coherence $\mathcal{C}_{cc}$, we apply a unitary
transformation to the reduced density matrices $\rho_{A}(T)$ and $\rho_{B}(T)$. As a result,
the unitary matrix is expressed as 
\begin{equation}
U_{A,B}=\left(\begin{array}{cc}
\cos\theta_{A,B} & -e^{i\varphi_{A,B}}\sin\theta_{A,B}\\
e^{-i\varphi_{A,B}}\sin\theta_{A,B} & \cos\theta_{A,B}
\end{array}\right).\label{eq:7}
\end{equation}
Let us consider $\widetilde{\rho}_{AB}(T)=\widetilde{U}_{AB}\,\rho_{AB}(T)\,\widetilde{U}^{\dagger}_{AB}$, 
where $\widetilde{U}=U_{A}\otimes U_{B}$. This unitary transformation establishes the relationship between global coherence and local 
coherence for various choices of the parameters $\theta_{A}$, $\theta_{B}$, $\varphi_{A}$
and $\varphi_{B}$. By setting $\varphi_{A}=0$, $\varphi_{B}=0$ and
\begin{equation}
\mathcal{\theta}_{A}=\arctan \left[\frac{\chi_{A}+\sqrt{\chi_{A}^{2}+4(\rho_{13}+\rho_{24})^{2}}}{2(\rho_{13}+\rho_{24})}\right].
\end{equation}
\begin{equation}
\mathcal{\theta}_{B}=\arctan \left[\frac{\chi_{B}+\sqrt{\chi_{B}^{2}+4(\rho_{12}+\rho_{34})^{2}}}{2(\rho_{12}+\rho_{34})}\right].
\end{equation}
Where $\chi_{A,B}=\rho_{11}\pm\rho_{22}\mp\rho_{33}-\rho_{44}$, into Eq. (\ref{eq:7}), we obtain a matrix that diagonalizes $\rho_{A}(T)$ and $\rho_{B}(T)$. This step provides us the basis set in which subsystems $A$ and  $B$ are locally incoherent. Finally, by substituting Eq.(\ref{eq:7}) into the Eq.(\ref{eq:6}), we derive an explicit expression for the correlated coherence.


\section{Results and Discussions}


\subsection{Population}

\begin{figure}
\includegraphics[scale=0.23]{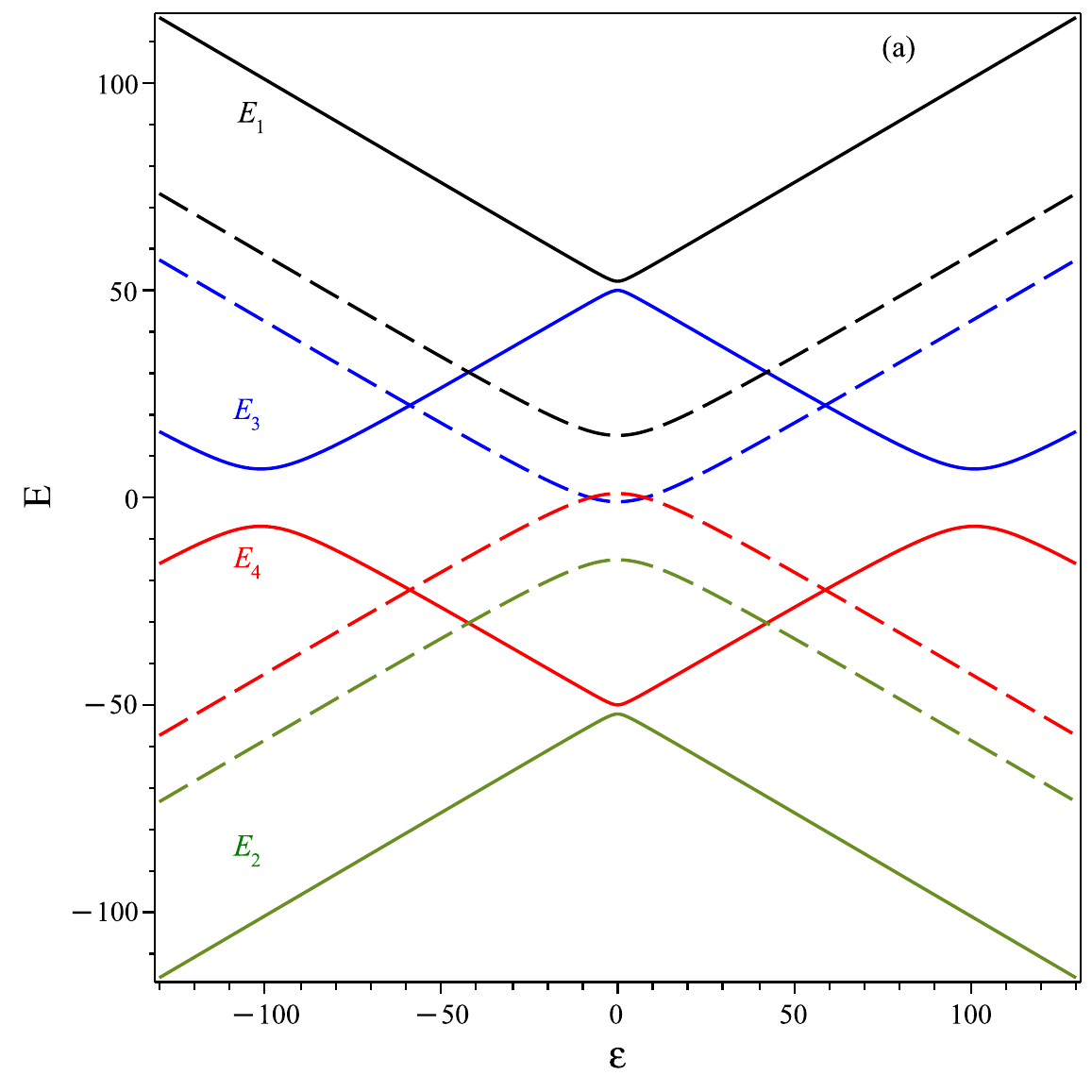}\includegraphics[scale=0.23]{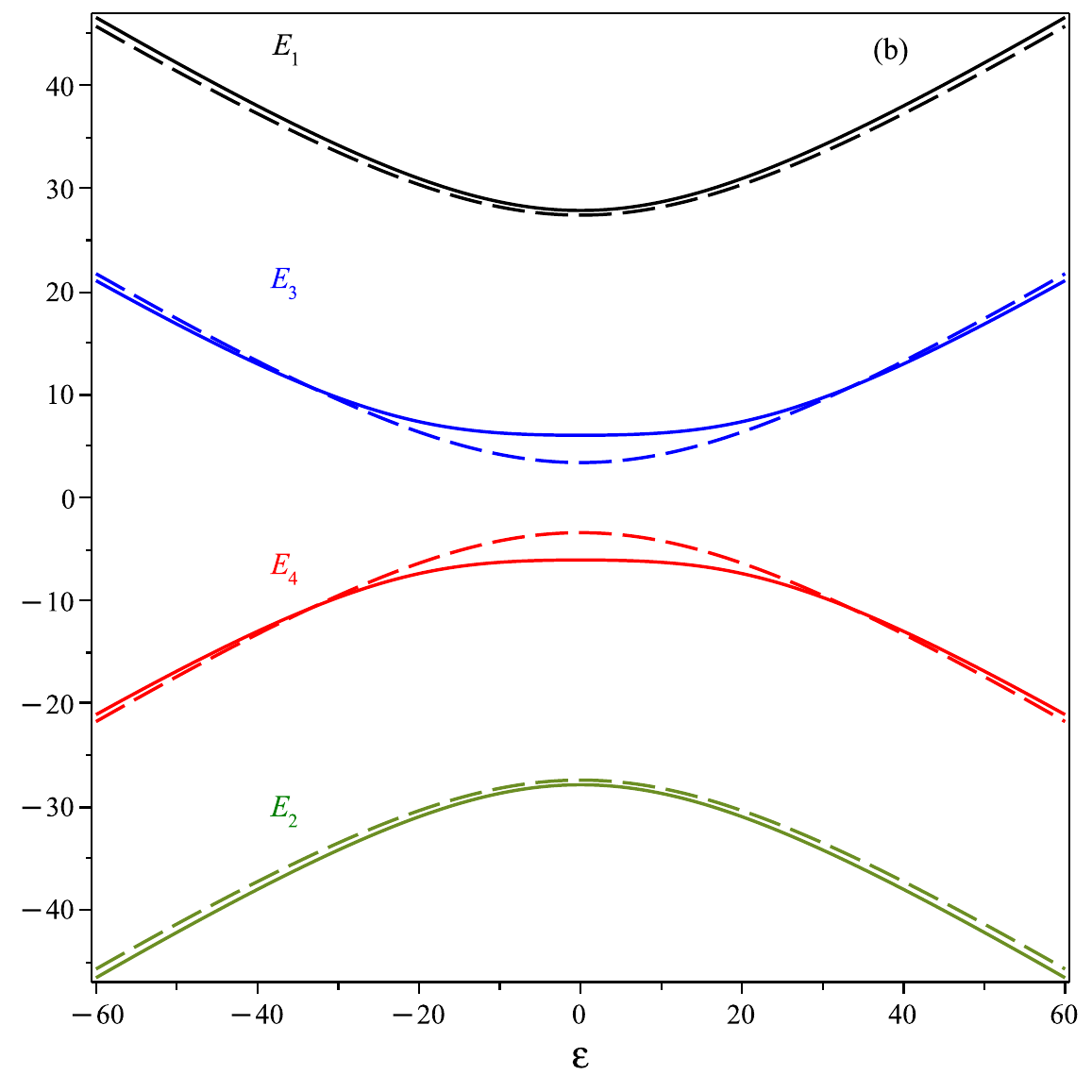}
\caption{\label{fig:energy} Energy spectrum of the DQD Hamiltonian $H$ as
a function of $\varepsilon$. (a) For $t=7$, $B_{z}=16$, with $B_{x}=0$(dashed curves)  and $B_{x}=100$(solid curves). (b) For $t=15.4$, $B_{z}=24$, again comparing $B_{x}=0$(dashed curves)  and $B_{x}=10$(solid curves).}
\end{figure}

Firstly, we investigate the influence of the tunneling coefficient
$t$ and magnetic field on the energy levels at zero temperature.
In Fig. \ref{fig:energy}, we have plotted  the energy levels versus 
inter-dot detunning $\varepsilon$.  
In Fig. 2(a), we plot the energy spectrum for $t=7$, $B_{z}=16$. The energy 
levels for $B_{x}=0$ are indicated by dashed lines, while the solid lines 
show the energy levels for a very intense transverse magnetic field $B_{x}=100$, resulting in two anticrossing points at  $\varepsilon=0$, specifically between energy levels $E_{2}$ and $E_{4}$, as well as between  $E_{1}$ and $E_{3}$. Additionally, the figure shows anticrossing at $\varepsilon\approx \pm101.9$ for energy levels $E_{3}$ and $E_{4}$. From the above analysis, it is evident that the transverse magnetic field contributes significantly to the formation of anticrossing points in the electron energy spectrum.  Fig. 2(b) illustrates a scenario for $t=15.4$, $B_{z}=24$ with a transverse magnetic field $B_{x}=10$ \cite{benito}. We can easily observed that the field in the $x$ direction increases the 
anticrossings at $\varepsilon=0$.

\begin{figure}
\includegraphics[scale=0.40]{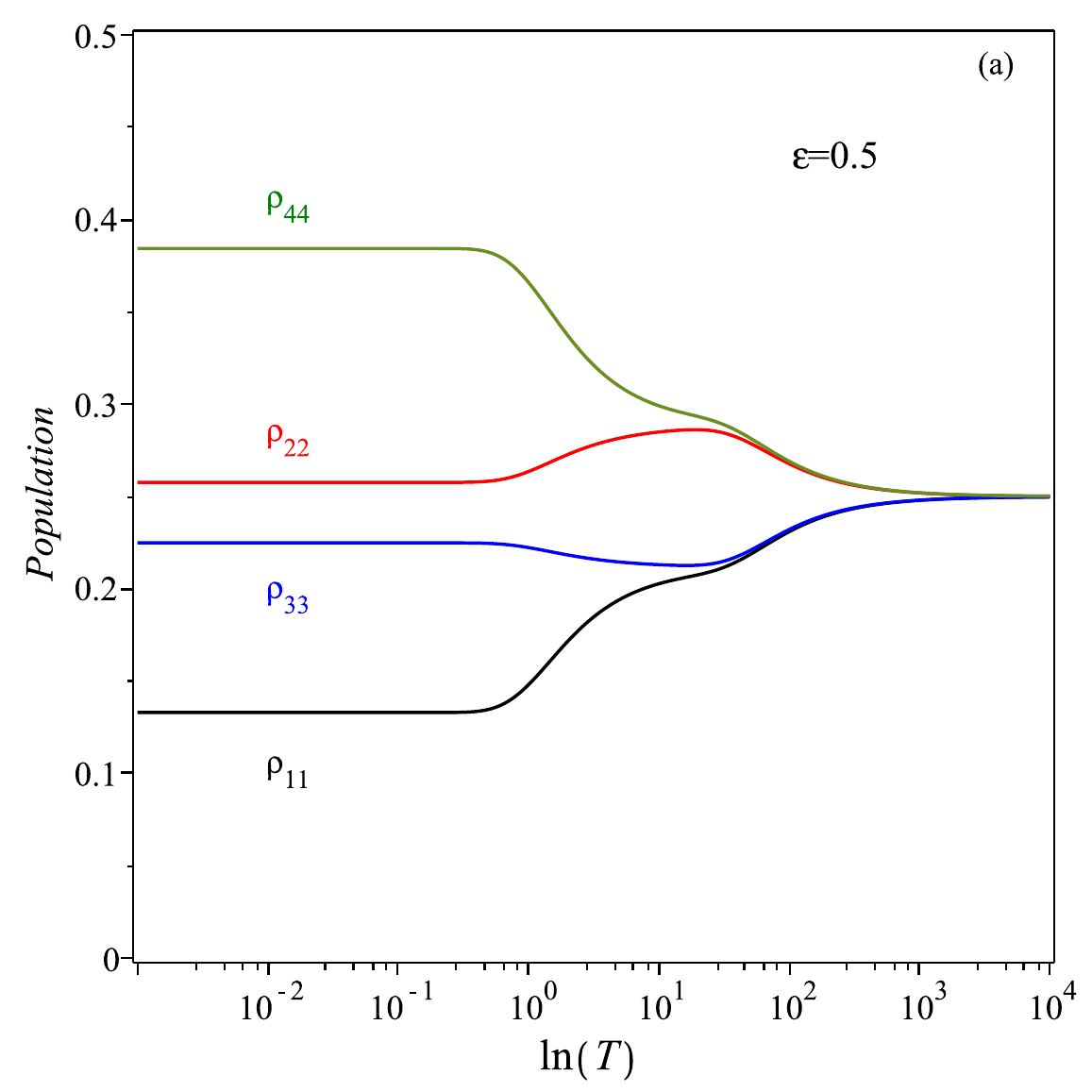}
\includegraphics[scale=0.40]{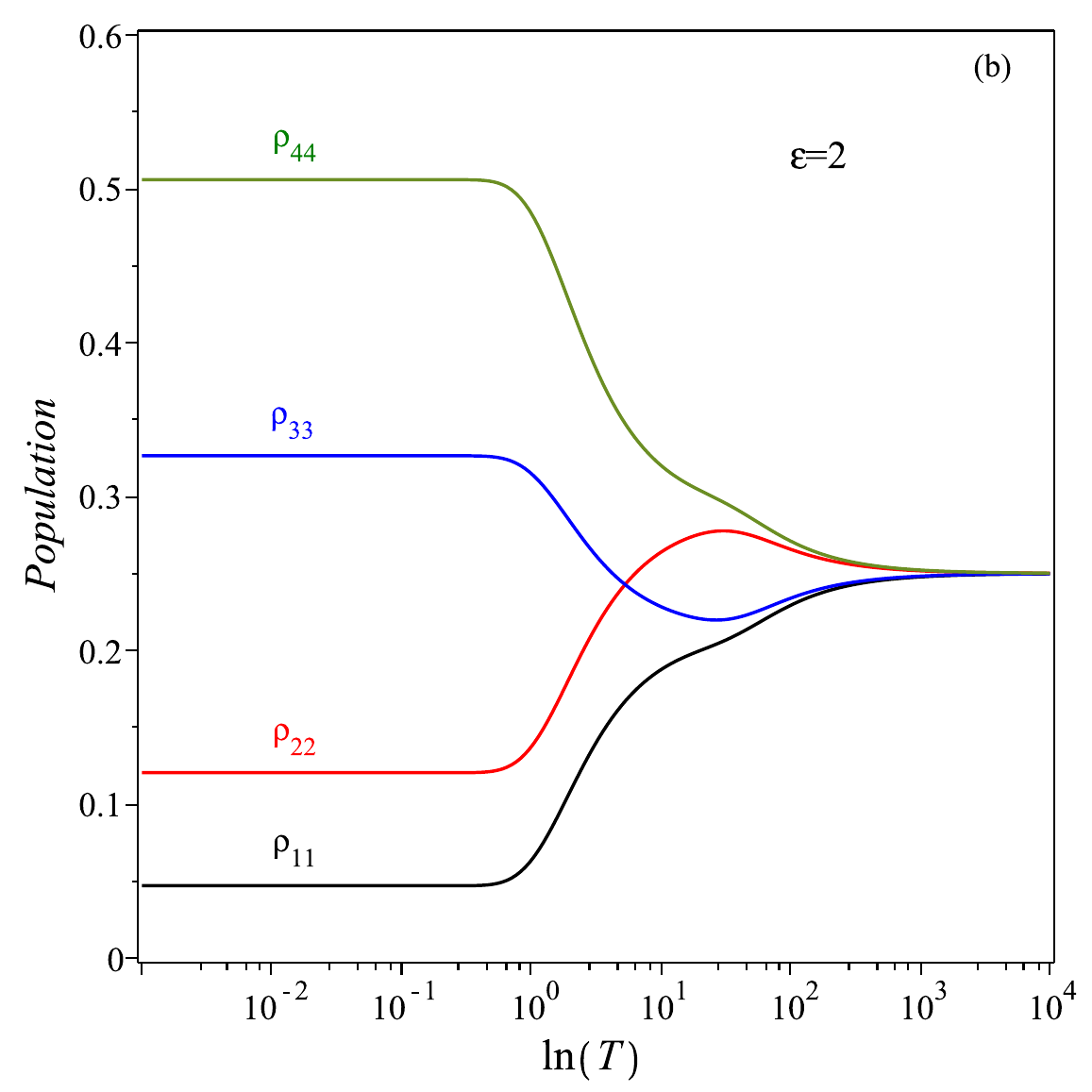}\caption{\label{fig:popu} Populations as a function of temperature $T$, displayed on a logarithmic scale, are shown for fixed parameters $t=7$, $B_{z}=16$, and $B_{x}=100$. (a)  $\varepsilon=0.5$. (b) $\varepsilon=2$.}
\end{figure}

Fig. \ref{fig:popu} illustrates the variation of population with respect to 
temperature $T$ on a logarithmic scale, for fixed values of $t=7$, $B_{z}=16$, $B_{x}=100$, with two different values of the inter-dot detuning 
parameter $\varepsilon=0.5$ and $\varepsilon=2$.  Fig.  \ref{fig:popu}(a) 
depicts the population under conditions of weak inter-dot 
detuning ($\varepsilon=0.5$). At low temperatures, the populations of the 
system remain constant. However, as the temperature increases, 
thermal fluctuations cause the populations of states $\rho_{33}$ 
and $\rho_{44}$ to decrease, while the populations of states $\rho_{11}$ 
and $\rho_{22}$ increase. In this scenario, there is no population inversion. 
Anyway, at higher temperatures, the populations are equally distributed, reaching 
the value 0.25.  Fig.  \ref{fig:popu}(b) displays population plots for inter-dot 
detuning ($\varepsilon=2$). Notably,  in the $T=0$ scenario, $\rho_{33}$ has 
the second-highest population, in contrast to the previous scenario, due to the 
significant inter-dot detuning value.  At low temperatures, the populations remain constant. However, as the temperature increases, a population inversion occurs between $\rho_{33}$ and $\rho_{22}$. Subsequently, at high temperatures, the populations become equally distributed.



\subsection{Concurrence}

First, we examine how the concurrence $\mathcal{C}$ is affected by the
temperature $T$, the transverse magnetic field $B_{x}$, and the interdot energy detuning $\varepsilon$. Next, we investigate the thermal fidelity of the model.

\begin{figure}
\includegraphics[scale=0.64]{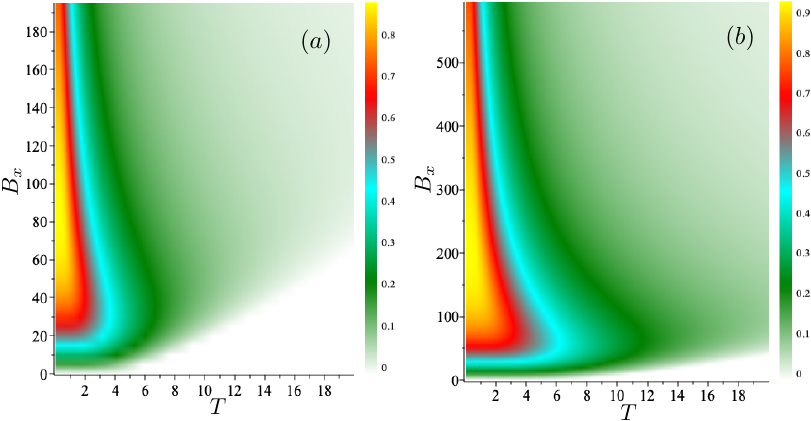}\caption{\label{fig:C-T} Density plot of the concurrence $\mathcal{C}$ as a function of the transverse magnetic field $B_{x}$ and temperature $T$. The parameters used for the plot are  $\varepsilon=1$,  (a) $t=7$,  the longitudinal magnetic field $B_{z}=16$.  (b) $t=15.4$,  the longitudinal magnetic field $B_{z}=24$.}
\end{figure}

In Fig. \ref{fig:C-T}, we illustrate the density plots of the concurrence $\mathcal{C}$ as a function of the transverse magnetic field $B_{x}$ and temperature $T$, for two different parameter sets. In Fig. \ref{fig:C-T} (a) for fixed values of  $\varepsilon=1$, $t=7$, and the longitudinal magnetic field $B_{z}=16$. The plot reveals 
two distinct entanglement regimes.  Initially,  the concurrence $\mathcal{C}$ is low for small values of the transverse magnetic field $B_{x}$, particularly in the range of $B_{x}\approx0$ to $B_{x}\approx20$. However, as $B_{x}$ increases, the concurrence
 also rises, reaching peak values in the range of approximately  $B_{x}$ between $60$ and $100$. Beyond this interval, with  $B_{x}$ values exceeding $100$, the concurrence decreases again. Regarding temperature, it is observed that entanglement is more significant at low temperatures (approximately between $T\approx0$ to $T\approx5$), gradually decreasing as the temperature rises. This behavior is expected, as the increase in temperature introduces thermal fluctuations, which tend to reduce quantum coherence and, consequently, the degree of entanglement in the system.
The Fig  \ref{fig:C-T}(b), with $t=15.4$ and $B_{z}=24$ fixed, the region with low transverse magnetic field values exhibits weak concurrence. However, within the range of $B_{x}$ between 100 to 200, the concurrence reaches its peak, as indicate by the yellow areas. This suggests that entanglement is enhanced at moderate $B_{x}$ values before decreasing again at even higher $B_{x}$ values. Similar to what is observed in Fig  \ref{fig:C-T}(a), entanglement is more robust at low temperature, with $T$ approximately between 0 to 8. However, as the temperature rises, entanglement gradually decays, albeit more slowly.

\begin{figure}
\includegraphics[scale=0.64]{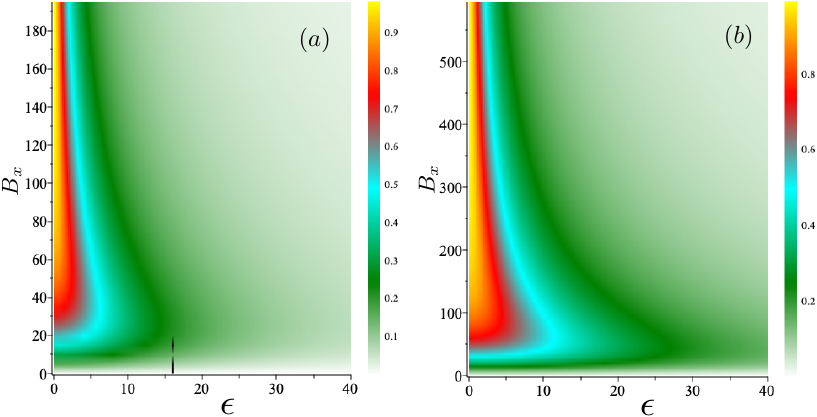}
\caption{\label{fig:C1} Density plot of the concurrence $\mathcal{C}$ as a function of the transverse magnetic field $B_{x}$ and the energy detuning $\varepsilon$ , with $T=0.2$, (a) tunneling parameter $t=7$, and the longitudinal magnetic field $B_{z}=16$. (b) tunneling parameter $t=15.4$, and the longitudinal magnetic field $B_{z}=24$.}
\end{figure}

Fig. \ref{fig:C1} presents density plots of concurrence $\mathcal{C}$ as a function of the transverse magnetic field $B_{x}$ and energy detuning $\varepsilon$. The temperature is kept constant at $T=0.2$. In Fig.  \ref{fig:C1} (a), with $t=7$ and $B_{z}=16$, it is observed that entanglement is initially weak for low values of the transverse magnetic field $B_{x}$ and arbitrary values of the energy detuning $(\varepsilon)$, as indicated by the green and blue regions in the figure.  However, for low values of $\varepsilon$ and as $B_{x}$ increases beyond this range, concurrence steadily rises, reaching its maximum entanglement level. This behavior highlights the significant sensitivity of entanglement to variations in both $B_{x}$ and $\varepsilon$. Regarding the dependence on $\varepsilon$, it becomes evident that  the system exhibits a high level of entanglement for small energy detuning values. On the other hand, as  $\varepsilon$ increases, it is observed that  for $B_{x}\approx 30$,  concurrence reaches a peak value of approximately $\mathcal{C}=0.6$ when  $\varepsilon$ is close to 5.  Beyond this point, entanglement decreases sharply,  becoming notably weak for higher detuning values. In Fig. \ref{fig:C1} (b), with $t=15.4$ and $B_{z}=24$ \cite{benito}, a similar trend to that observed in Fig. \ref{fig:C1} (a) is noted regarding the response of concurrence $\mathcal{C}$ to increases in $B_{x}$ and $\varepsilon$. However, the range of  $B_{x}$ where entanglement reaches its maximum values expands significantly, starting  around  $B_x \approx 200$. This expansion indicates that a stronger longitudinal field $(B_{z})$ enables robust entanglement only at high $B_{x}$ values. In all cases. it is observed that the maximum entanglement is achieved for null or low detuning energy values, while the transverse magnetic field plays a crucial role in the emergence of entanglement.

\begin{figure}
\includegraphics[scale=0.44]{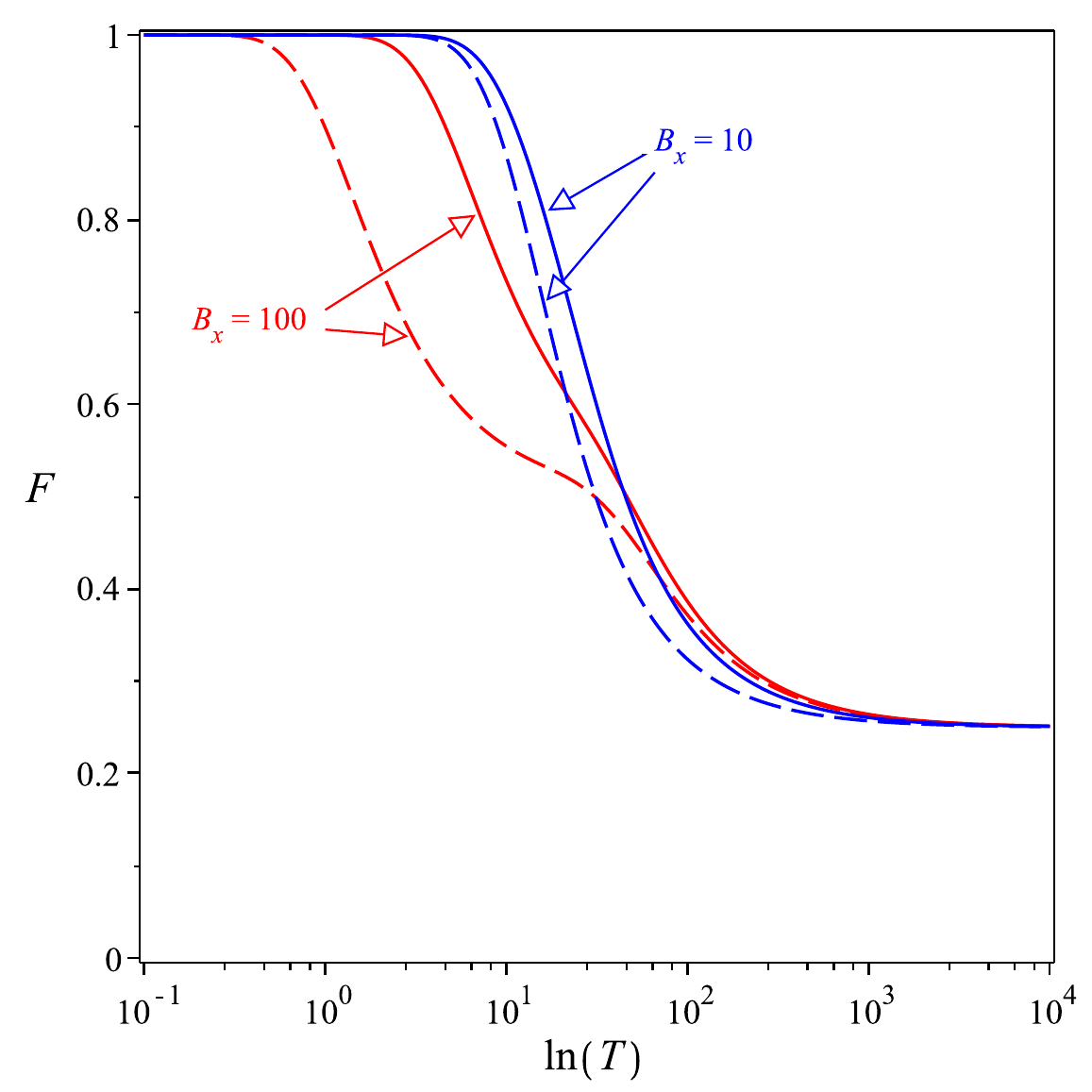}\caption{\label{fig:Fidelity} The fidelity $F$ is shown as a function of temperature $T$ on a logarithmic scale. For the parameters $t=7$, $B_{z}=16$, and $B_{x}=100$, we have $\varepsilon=0$ (dashed red curve) and $\varepsilon=10$ (solid red curve). For $t=15.4$, $B_{z}=24$, and $B_{x}=10$, we have $\varepsilon=0$ (dashed blue curve) and $\varepsilon=50$ (solid blue curve).}
\end{figure}

In Fig. \ref{fig:Fidelity}, the fidelity $F$ between the ground state  $|\varphi_{2}\rangle$ and the thermal state $\rho_{AB}(T)$ is plotted as a function of temperature $T$ on a logarithmic scale for two sets of parameters. As the temperature approaches zero, the fidelity of the mixed state converges to the ground-state fidelity, reaching $F=1$. However, as the temperature rises, the ground state begins to mix with the excited states, causing the fidelity to decrease steadily. Moreover, it is observed that for nonzero energy detuning $\varepsilon$ (solid curves), the fidelity remains higher compared to the case where $\varepsilon=0$ (dashed curves). This demonstrates that the fidelity is more robust for nonzero values of $\varepsilon$ at low temperatures, reflecting greater system stability in this regime. As the temperature continues to rise, however, the fidelity becomes nearly independent of $\varepsilon$, while the system's robustness is significantly compromised due to thermal fluctuations at higher temperatures. 

\subsection{Thermal quantum coherence}


To estimate the quantum correlations stored internally in quantum correlations we will use correlated coherence.

\begin{figure}
\includegraphics[scale=0.44]{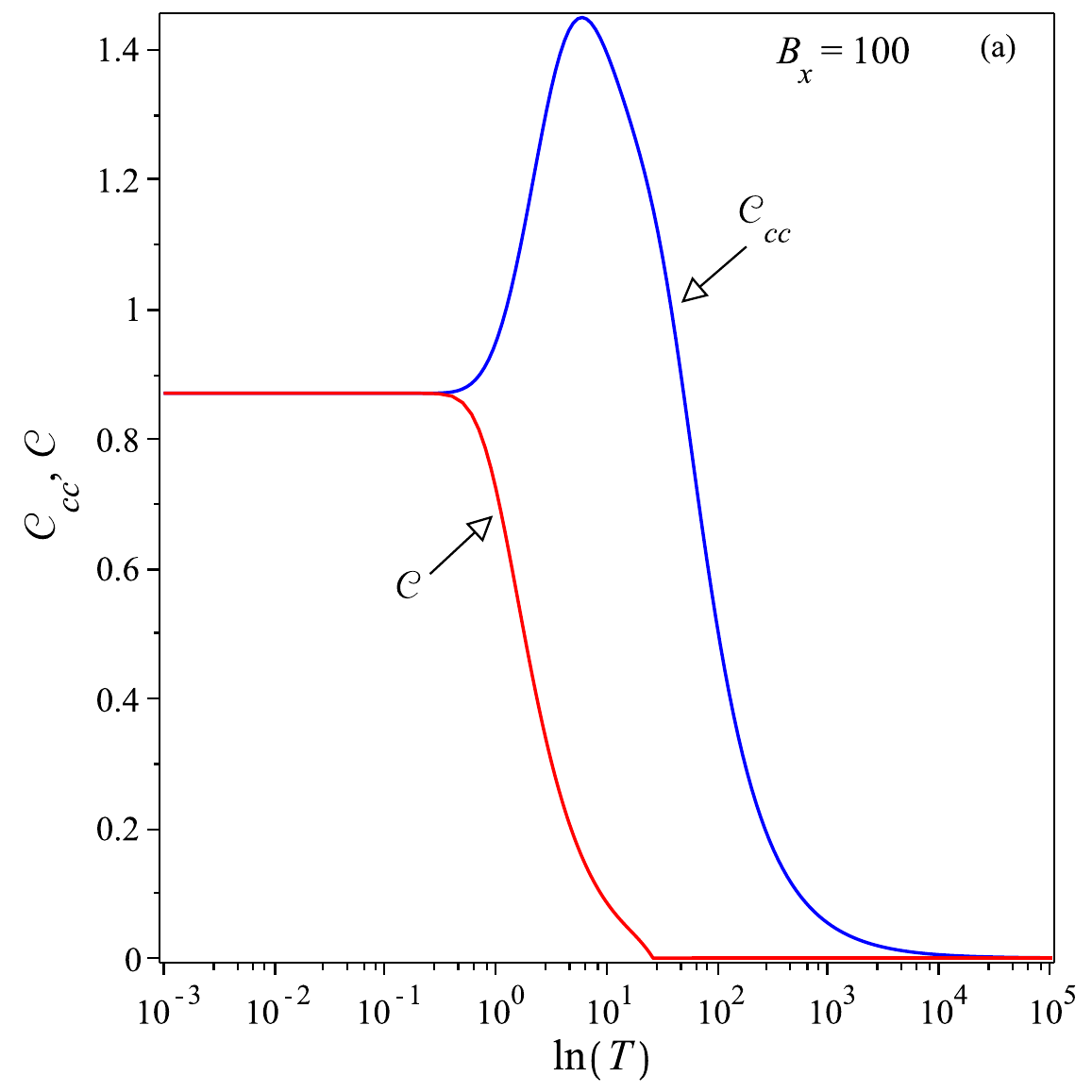}
\includegraphics[scale=0.44]{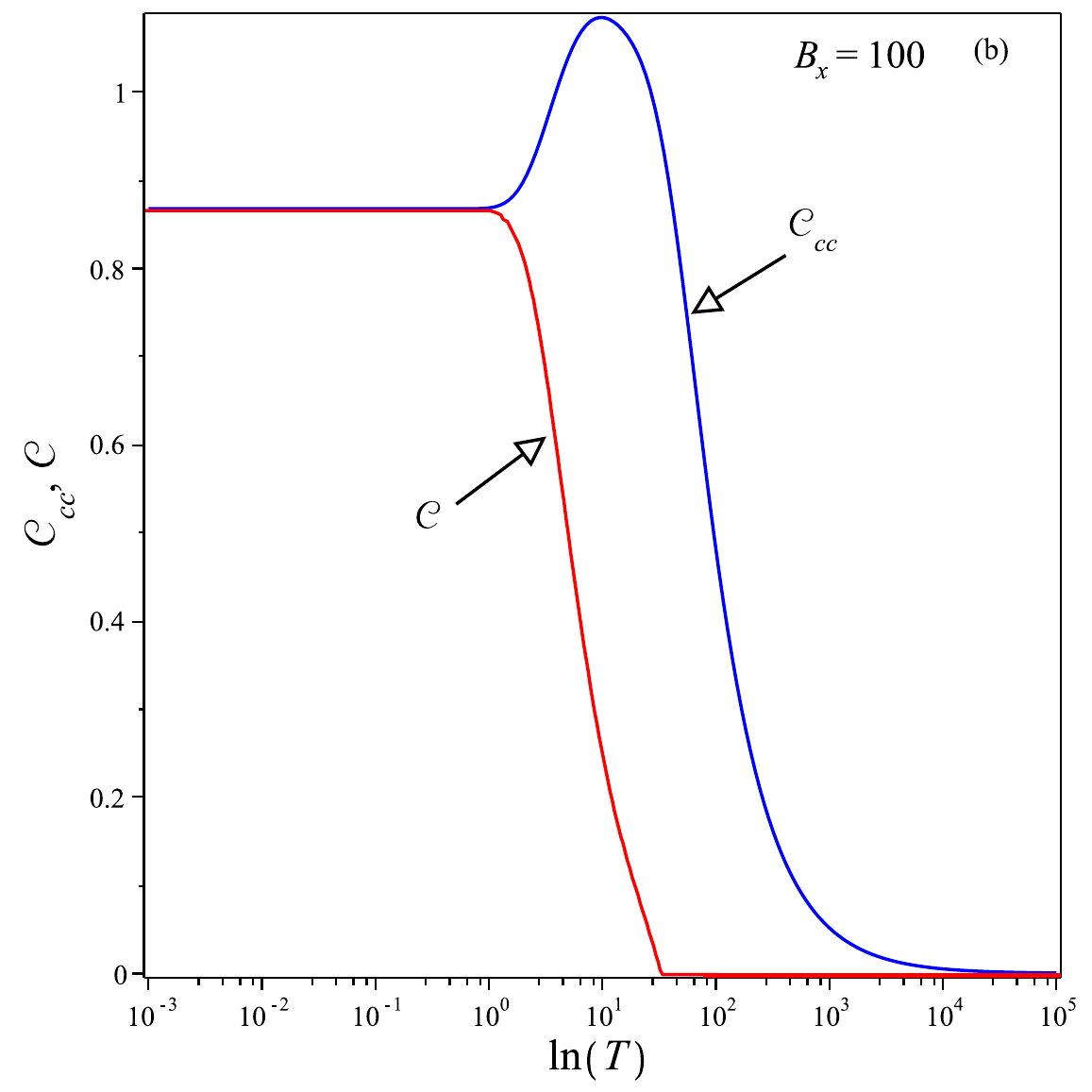}
\caption{\label{fig:7} Correlated coherence $\mathcal{C}_{cc}$ (blue curve) and Concurrence $\mathcal{C}$ (red curve) as a functions of temperature $T$, displayed on a logarithmic scale, with $\varepsilon=1$ and $B_{x}=100$. (a) $t=7$, and $B_{z}=16$. (b) $t=15.4$, and $B_{z}=24$. }
\end{figure}


In Fig. \ref{fig:7}, we present the plots of both $\mathcal{C}_{cc}$ (correlated coherence) and $\mathcal{C}$ (concurrence) as 
a function of temperature $T$ on a logarithmic scale, with the detuning energy $\varepsilon=1$ and $B_{x}=100$. Here, correlated coherence is evaluated in 
the incoherent basis, with the local coherence parameters $\theta$ specified by Eqs.  (\ref{eq:rha}) and (\ref{eq:rhb}), while $\varphi=0$. 
In Fig. \ref{fig:7} (a), we consider the parameters $t=7$ and $B_{z}=16$. In this figure is observed that as $T$ increases 
from low values, concurrence $C$ initially remains stable but then rapidly decreases once $T$ reaches a certain threshold. As the 
temperature increases, the entanglement decreases sharply to nearly zero as $T$ reaches the intermediate range, suggesting that 
entanglement cannot withstand thermal fluctuations at high temperatures. In contrast, the behavior of the correlated 
coherence $\mathcal{C}_{cc}$ reveals greater subtlety. At low temperatures, the system's entangled quantum correlations are entirely contained within the quantum coherence, suggesting that the correlated coherence fully captures all thermal entanglement information in this regime.  As the temperature increases, thermal fluctuations cause a slight rise in quantum coherence, while entanglement gradually diminishes. As the temperature continues rise, coherence eventually vanishes upon reaching a threshold temperature. In Fig. \ref{fig:7} (b), with parameters $t=15.4$ and $B_{z}=24$, the behaviors of $\mathcal{C}_{cc}$ and $\mathcal{C}$ exhibit qualitative similarities to those in Fig.  \ref{fig:7} (a), but with some significant differences. At low temperatures, both quantities show high values, indicating that quantum correlations and correlated coherence remain robust. However, the separation between the two curves occurs at a higher temperature range, suggesting greater thermal resilience of the system under these conditions. Furthermore, the peak of  $\mathcal{C}_{cc}$  shifts to $T\approx9.8$, compared to $T\approx6.01$ in Fig. \ref{fig:7} (a), which can be attributed to the increase en $B_{z}$ that modifies the system's dynamics and enhances correlated coherence at higher temperatures. The concurrence $\mathcal{C}$, in turn, displays a continuous decline with increasing temperature, similar to Fig. \ref{fig:7} (a), but with a steeper drop at intermediate temperatures, indicating greater vulnerability of quantum correlations in this regime. These results highlight the critical role of the parameters $t$ and $B_{z}$ in defining the systems the system's thermal robustness, directly influencing the response of correlated coherence and the preservation of quantum correlations under different thermal conditions.

\section{Conclusions}

This paper considers a device composed of a single electron confined in a double
quantum dot, where the electron's spin is subject to a homogeneous magnetic field $B_{z}$ and a magnetic field gradient $B_{x}$ within a thermal bath. Our primary objective was to solve the model exactly and investigate the effects of temperature on quantum coherence. Our results indicate that the energy spectrum exhibits an anticrossing phenomenon drive by the transverse magnetic field. The model shows that thermal entanglement for a single electron can be achieved through the interplay between charge and spin qubits. 
We further explore the impact of the transverse magnetic field on population dynamics and concurrence. The results indicate that both are highly sensitive to temperature and the strength of the transverse magnetic field. Specifically, for weak magnetic fields, entanglement remains weak across all detuning energies. In contrast, for strong magnetic fields, concurrence becomes pronounced, particularly at low detuning energies. Regarding temperature, concurrence stabilizes at a plateau at low temperatures; however, as temperature increases, it diminishes due to the increasing impact of thermal fluctuations. Moreover, we identified a direct relationship between entanglement and quantum coherence. Although quantum coherence is inherently basis-dependent, it is possible to select an incoherent basis for local coherence, enabling the determination of correlated coherence. Our findings indicate that, at low temperatures, the correlated coherence is equal to the concurrence.
Additionally, as the temperature increases, the model revealed an unusual thermally-induced increase in correlated coherence, driven by the emergence of non-entangled quantum correlations as entanglement decreases. At sufficiently high temperatures, quantum entanglement vanishes entirely, as thermal fluctuations become the dominant influence on the system's behavior.
\section{Acknowledgments}

This work was partially supported by CNPq, CAPES and Fapemig. M. Rojas
would like to thank National Council for Scientific and Technological Development-CNPq grant 317324/2021-7 and Fapemig Grant  APQ-02226-22.

\end{document}